\documentclass{jaa}

\usepackage{graphicx}

\newcommand{\gcc}{\mbox{ g cm}^{-3}}
  \usepackage{aas_macros}
  \usepackage{natbib}

\begin{document}

\title{Revisiting field burial by accretion onto neutron stars}


\author{Dipanjan Mukherjee \textsuperscript{1*}}
\affilOne{\textsuperscript{1}Research School of Astronomy and Astrophysics, Australian National University, Canberra, ACT 2611, Australia\\}


\twocolumn[{

\maketitle

\corres{dipanjan.mukherjee@anu.edu.au}

\msinfo{--}{--}{--}

\begin{abstract}
The surface magnetic field strength of millisecond pulsars (MSPs) is found to be about 4 orders of magnitude lower than that of garden variety radio pulsars (with a spin of $\sim 0.5-5$ s and $B\sim 10^{12}$G). The exact mechanism of the apparent reduction of field strength in MSPs is still a subject of debate. One of the proposed mechanisms is burial of the surface magnetic field under matter accreted from a companion. In this article we review the recent work on magnetic confinement of accreted matter on neutron stars poles. We present the solutions of the magneto-static equations with a more accurate equation of state of the magnetically confined plasma and discuss its implications for the field burial mechanism.

\end{abstract}


}]


\doinum{12.3456/s78910-011-012-3}
\artcitid{\#\#\#\#}
\volnum{123}
\year{2016}
\pgrange{23--25}
\setcounter{page}{23}
\lp{25}

\section{Introduction}
The apparent surface magnetic field strength of millisecond pulsars ($\sim 10^8$G) is much lower than that of normal radio pulsars ($\sim 10^{12}$G). The cause of the lower field strength of millisecond pulsars (MSPs) has long been a subject of debate in the literature. With the discovery of pulsars with high magnetic fields to have ages $\sim 10^8-10^9$ years \citep[][ and others]{kulka86,calla89,koest92}, spontaneous exponential decay of the magnetic field was ruled out, as it predicts a short turnover time of a few million years.  Other works involving pulsar population studies \citep{baile89,bhatt92,fauch06} have also ruled out spontaneous field decay in isolated pulsars. 

The discovery of fast spinning pulsars in binary systems \citep[e.g. PSR 1913+16][]{hulse75} had led several authors \citep[e.g. ][]{smarr76,srini82,radha84,alpar82} to propose a new evolutionary pathway of recycled pulsars to explain the observed fast spins of MSPs. In such a scenario, neutron stars are spun up to periods of a few  milliseconds, by transfer of angular momentum due to accretion of matter from its companion. Such a ``recycling scenario" was further supported later by the discovery of on-going accretion in the millisecond X-ray pulsars SAX~J1808.4--3658 \citep{wijna98}. 

The success of the accretion induced recycling scenario in explaining the spin evolution of MSPs also strongly suggests an
accretion induced evolution of the magnetic field. One of the earliest of such suggestions was by \citet{bisno74},
even before the discovery of MSPs. From subsequent works a few prominent mechanisms have emerged as likely explanations
of the apparent reduction in surface field strengths in MSPs: 
\begin{itemize}
\item burial of surface magnetic field by piled up accreted matter \citep{romani90,cumming01,melatos01,choudhuri02,konar04,melatos04,payne07}, 
\item accretion induced enhancement of ohmic decay of crustal magnetic fields \citep{konar97,konar99a}, 
\item flux expulsion from superconducting core and its subsequent decay in the crust \citep{srini90,jahan94,konar99b}, 
\item thermo-magnetic evolution of crustal field \citep{blond86} and 
\item rotation induced re-orientation of the magnetic field due to crustal tectonic motions \citep{ruder91a,ruder91b}.
\end{itemize}

Of the above, the field burial mechanism has received prominence, due to the physically motivated modelling pursued in recent works \citep{melatos01,melatos04,payne07,priymak11}. According to this conjecture, the accreted matter after being channelled to the poles spreads equator-wards, dragging the magnetic field lines with it \citep[see Fig. 6 of][ for an illustrative diagram]{priymak11}. The large scale deformation of the magnetic fields, stretched from the poles to the equator, will result in large local screening currents. Compression from subsequent accretion will bury the deformed field into deeper layers of the crust, reducing the apparent external dipole moment. Although promising, there are doubts whether such large scale deformation of the magnetic field topology can be sustained in the presence of MHD instabilities \citep{bhatta99,litwin01,cumming01,mukhe13d}. MHD instabilities, if present, will operate on local dynamical time scales ($\sim 10^{-3}$ s), much shorter than that of the long term accretion time scales required for the burial process to operate. 

In this article we will review the existing work on the magnetic confinement of accreted matter on neutron stars, as well as present new results with an updated equation of state. We then discuss the implications of the results for the field burial scenario. The outline of this article is as follows. Section~\ref{sec.mountains} discusses the formulation of the Grad-Shafranov equation and its solution describing the structure of the magnetically confined accretion mound. The solutions are presented for a new equation of state of the plasma, which is more accurate in representing the plasma state over a wide range of densities. We review in detail the differences between the approaches adopted by previous authors and its implications on the resultant solutions. In Sec.~\ref{sec.instabilities} we review the various dominant MHD instabilities that can destabilise the confined mound, severely restricting the efficiency of field burial. The results are finally summarised in Sec.~\ref{sec.summary} with a discussion on the future directions in Sec.~\ref{sec.future}

\section{Forming magnetically confined mounds}\label{sec.mountains}
\subsection{Equation of state of the confined plasma}
The choice of equation of state of the plasma significantly affects the mass and size of the confined mound \citep{priymak11}. The state of the plasma and its pressure is determined by its  density and temperature. For a plasma with temperature $T\!<\!T_F\!=\!\frac{m_ec^2}{k_B}\left[\!\sqrt{x_F^2\!+\!1}-1\right]$ ($x_F=p_F/(m_ec)$, $p_F$ being the Fermi momentum and $m_e$ the electron mass), the dominant contribution to the total pressure is from the degeneracy pressure from an electron Fermi gas, given by: 
\begin{align}
p&=\frac{\pi}{3}\frac{m_e^4 c^5}{h^3}\left[x_F(x_F^2+1)^{1/2}(2x_F^2-3) \right. \nonumber\\
 &\left. +3\ln \left(x_F+\sqrt{1+x_F^2}\right)\right] \label{eq.fermip},\\
\mbox{ with }x_F \!\!&=\!\! \frac{1}{m_e c}\left(\frac{3 h^3}{8 \pi \mu _em_p}\right)^{1/3} \! \rho ^{1/3}. \label{eq.fermiM} 
\end{align}
For a temperature of $T \sim 2\times10^7$K, typical of the hotspots in HMXBs \citep{coburn02}, the plasma is degenerate for $\rho \gtrsim 10^3 \mbox{ g cm}^{-3}$. Since the densities at the base of the mound are as high as $\sim 10^8-10^9 \gcc$, a degenerate equation of state (hereafter EOS)  is the apt choice. 

\begin{figure}
	\centering
        \includegraphics[width = 8cm,keepaspectratio] {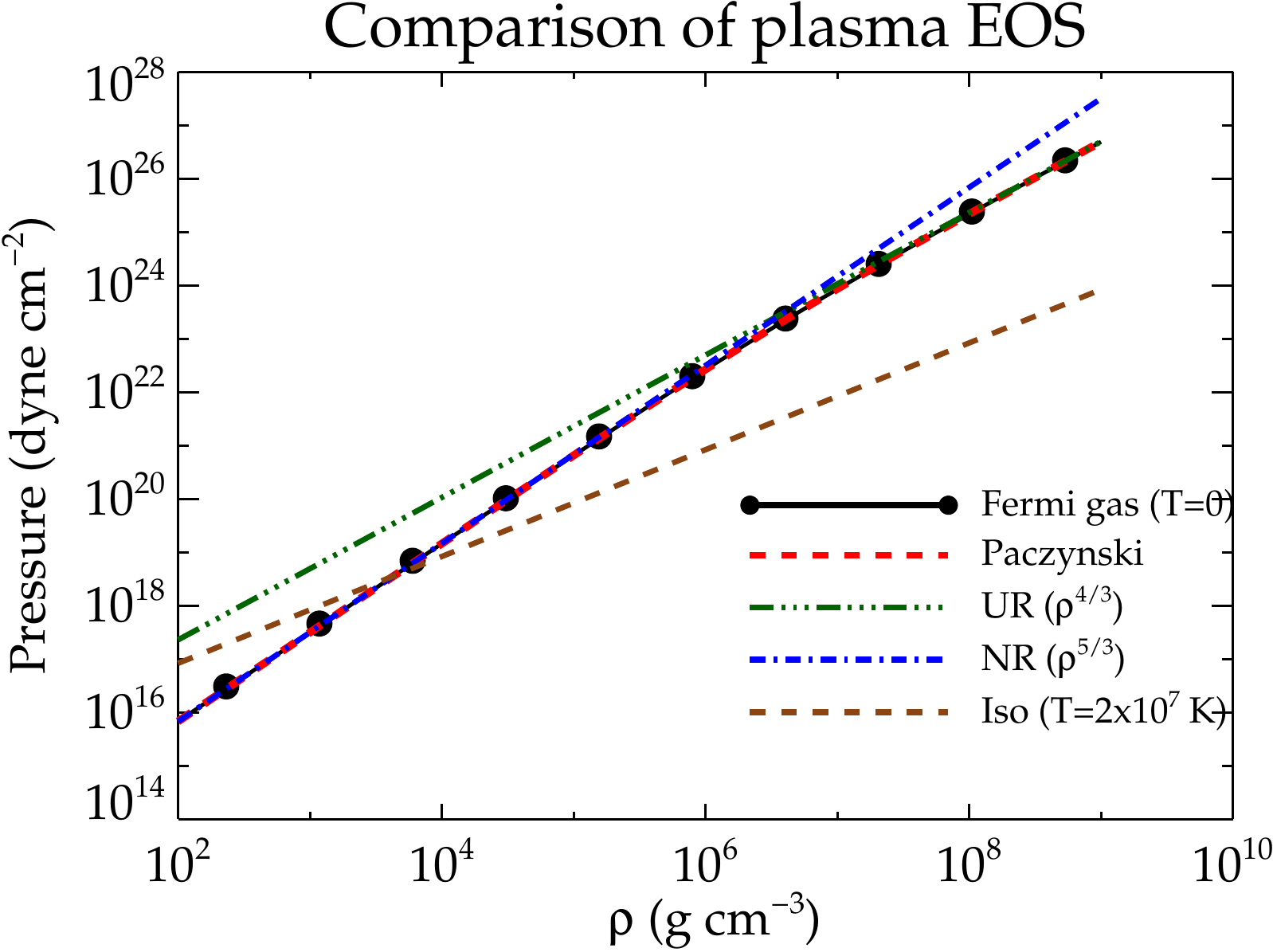}
	\caption{ \small Comparison of the pressure-density relation for different plasma states: a) T=0 degenerate Fermi gas (solid black line with dots), b) the Paczynski approximation to the Fermi gas as in eq.~\ref{eq.pacz} (red dashed) c) ultra-relativistic degenerate gas with $p=4.89\times10^{14}\rho^{4/3} \mbox{ dynes cm}^{-2}$ (green with dash tripple dot) d) non-relativistic degenerate gas with $p=3.12\times10^{12}\rho^{4/3} \mbox{ dynes cm}^{-2}$ e) isothermal EOS with $T=2\times 10^7$ K$\sim 2$ keV (brown dashed). The Paczynski EOS  being a very close approximation, is indistinguishable from the Fermi EOS.}\label{fig.eos}
\end{figure}

Previously, several works have considered a classical ideal gas with an isothermal non-degenerate plasma while modelling the confined matter \citep{melatos04,payne07,vigelius08,vigelius09}. However, for densities interior to the mound, an isothermal EOS underestimates the pressure by several orders of magnitude from that of a realistic degenerate gas (see Fig.~\ref{fig.eos}). Hence, although favoured for its analytical ease, the results with an isothermal EOS are incorrect at the base of the mound where degeneracy pressure is expected to dominate.

Other works have considered a degenerate polytropic gas ($p \propto \rho ^\gamma$) with a single polytropic index: $\gamma = 4/3$ for an ultra-relativistic gas with $x_F \gg 1$ \citep{hameury83,melatos01,priymak11,dipanjan12} and $\gamma = 5/3$ for a non-relativistic approximation $x_F \ll 1$ \citep{priymak11,dipanjan12,priymak14}. However the approximate single polytrope EOS significantly overestimates the pressure for certain density ranges  as compared to the Fermi EOS (see Fig.~\ref{fig.eos}). A better approximation to the degenerate gas over a wide range of densities is given by \citet{pacz83} (hereafter the Paczynski EOS):
\begin{equation}
p=\frac{\pi}{3}\frac{m_e^4 c^5}{h^3} \frac{(8/5) x_F^5}{\left(1+(16/25)x_F^2\right)^{1/2}} \label{eq.pacz}
\end{equation}
which correctly asymptotes to the ultra and non-relativistic limits. The above is accurate to $\sim 1.8\%$ of the Fermi pressure (given in eq.~\ref{eq.fermip}). The simple analytic form of the Paczynski EOS and its high relative accuracy to the Fermi pressure ($\sim 1.8\%$) makes it a better choice for the semi-analytic modelling of the magnetically confined mound, as described in subsequent sections below.

\subsection{The Grad-Shafranov formulation with Paczynski EOS}\label{sec.pacz}
The structure of a magnetically confined accretion mound can be evaluated by solving the Euler equation in the 
static limit  under force balance.\begin{equation}
\nabla p + \rho \nabla \phi _g = \frac{(\nabla \times \boldsymbol{B}) \times \boldsymbol{B}}{4 \pi} \label{eq.eul} \\
\end{equation}
In this work we consider spherical coordinates ($r,\theta,\phi$), 
with the gravitational potential approximated by $\phi _g = (GM_s/R_s^2)r=gr$. Assuming axisymmetry ($\partial/\partial\phi=0$) and a poloidal magnetic field configuration ($B_\phi =0$), 
the magnetic field can be expressed in terms of the flux function $\psi$ as: 
\begin{equation}
\boldsymbol{B} = \frac{\nabla \psi (r,\theta) \times \hat{\phi}}{r \sin \theta}
\end{equation}
Thus the RHS of the Euler equation (eq.~\ref{eq.eul})  becomes
\begin{align}
\frac{(\nabla \times \boldsymbol{B}) \times \boldsymbol{B}}{4 \pi} &= \frac{-\Delta ^2 \psi}{4 \pi r^2 \sin ^2 \theta} \nabla \psi \label{eq.rhs}\\
\mbox{where }\Delta ^2 \psi &= \frac{\partial ^2 \psi}{\partial r^2} + \frac{\sin \theta}{r^2}\frac{\partial}{\partial \theta}
\left(\frac{1}{\sin\theta} \frac{\partial \psi}{\partial \theta}\right). \nonumber
\end{align}
Using the Paczynski EOS (eq.~\ref{eq.pacz}) and defining
\begin{equation}\label{eq.G}
G(x_F)=\frac{m_e c^2}{8 \mu _e m_p} \left[ \frac{15/2+(32/5)x_F^2}{\left(1+(16/25)x_F^2\right)^{1/2}} \right]
\end{equation}
we can express the LHS of the Euler equation as  
\begin{equation}
\nabla p + \rho \nabla \phi _g = \rho \nabla\left( G(x_F)+\phi _g \right) \label{eq.lhs}
\end{equation}
Eq.~\ref{eq.lhs} above assumes that the LHS can be expressed as a gradient. This is strictly true only for a barotropic equation of state where the pressure at any location is uniquely determined by the density alone. This assumes an isothermal plasma with a homogeneous composition, which for this work is assumed to be $\mu_e=2$ (ionised Helium). However nuclear reactions occurring in the deeper layers ($\rho \gtrsim 10^6 \gcc$) may result in variation in temperature and chemical composition inside the settling layers. Formulation of the Grad-Shafranov for such a non-barotropic plasma is non-trivial, as eq.~\ref{eq.lhs} can no longer be expressed as a gradient. See for example \citet{akgun13} for a semi-analytic modelling of magnetic equilibria for a non-barotropic star. Similar works on magnetically confined mounds have so far not been carried out. 

A barotropic approximation will be valid at times longer than the nuclear burning time scales \citep[typically hours to days, see ][]{brown98} where the incoming fuel is burnt and compressed as settling ash into the deeper layers to form a homogeneous mixture of unburnt hydrogen and helium and heavier elements. Hence the present results discussed will be valid while considering the long term settling of the accreted material after the nuclear reactions have been spent. See Sec.~\ref{sec.otherinst} for more on the effect of variation in chemical composition.

Combining eq.~\ref{eq.rhs} -- eq.~\ref{eq.lhs} the Euler equation (eq.~\ref{eq.eul}) can be written as 
\begin{equation}
\rho \nabla\left( G(x_F)+\phi _g \right) = \frac{-\Delta ^2 \psi}{4 \pi r^2 \sin ^2 \theta} \nabla \psi = \rho g\nabla r_0(\psi) \label{eq.eulfinal}
\end{equation}
In the RHS of eq.~\ref{eq.eulfinal}  we introduce the function $r_0(\psi)$ which depends only on the axisymmetric flux function $\psi$ \citep[e.g. see][]{litwin01,melatos04,dipanjan12}.
Evaluating eq.~\ref{eq.eulfinal} along a constant $\psi$ surface \citep[e.g.][]{melatos04}, we get the well known Grad-Shafranov equation \citep{shafr58}
\begin{equation}
\Delta ^2 \psi = -4 \pi r^2 \sin^2 \theta \, \rho g \frac{d r_0(\psi)}{d \psi} \label{eq.gs}
\end{equation}
with $r_0(\psi)$ being an unspecified function which defines the shape of the flux surfaces.

The distribution of the density in eq.~\ref{eq.gs} can be obtained by integrating eq.~\ref{eq.eulfinal} along a $\psi=$constant surface as
\begin{align}
\nabla \left( G(x_F)+\phi _g \right) &=  g\nabla r_0(\psi) \label{eq.cpsi1} \\
G(x_F) &= g(r_0(\psi)-r)+C(\psi). \label{eq.cpsi2}
\end{align}
with $C(\psi)$ being a constant of integration. $C(\psi)$ can be determined by assuming $r_0(\psi)$ to be the top of the mound i.e. $x_F\!=\!0\rightarrow \rho\!=\!0$ at $r=r_0(\psi)$.
Using the above criterion, eq.~\ref{eq.cpsi2} can be inverted to give the density distribution inside the accretion mound as:
\begin{align}
x_F^2 &= \frac{225}{512}\left(\eta ^2 -\frac{8}{3} + \eta \sqrt{\eta ^2 + \frac{16}{9}} \right) \label{eq.dens}\\
\mbox{ where } \eta &= \frac{2}{15}\frac{g \mu_e m_a}{m_ec^2}\left(r_0(\psi)-r\right)+1.
\end{align}
$m_e$ is the electron mass, $m_a$ the atomic mass unit and $\mu_e=2$  the mean molecular weight. The negative root in the inversion of eq.~\ref{eq.dens} has been discarded as it leads to $x_F^2 < 1$, implying unphysical densities. 

Numerically solving eq.~\ref{eq.gs} together with eq.~\ref{eq.dens} will determine the magnetic field and density structure of the magnetically confined accretion mounds \citep{melatos04,priymak11,dipanjan12}. Two different approaches have been taken in previous works while evaluating the solutions, with differences in assumptions of the boundary conditions and the function $r_0(\psi)$ leading to different results, as outlined below.

\subsection{Boundary Conditions}\label{sec.boundary}
 The choice of the boundary condition significantly affects the solutions of the 
Grad-Shafranov equation. Some authors \citep{melatos04,payne07,vigelius08,vigelius09,priymak11,priymak14} have assumed a free boundary condition ($\partial \psi/\partial r =0$). The advantage of assuming a free boundary is that the magnetic field can evolve from its initial guess value, which is desired to obtain reduction in the field strength as required in the burial scenario. However, since the magnetic field components are related to the flux function as
\begin{equation}
B_\theta = -\frac{1}{r \sin \theta}\frac{\partial \psi}{\partial r}\; ; \; B_r = \frac{1}{r^2 \sin \theta}\frac{\partial \psi}{\partial \theta},
\end{equation} 
such a boundary condition implies a radial magnetic field with $B_\theta =0$ at the boundaries. Hence assuming a free boundary comes at the cost of an unphysical monopole like field configuration. 

A physically consistent boundary condition must match with a force free vacuum solution outside the mound. In that regard, other works have assumed a fixed boundary at the edges of the compute domain, pinning the magnetic field value to that of a dipole (or its approximation) \citep[e.g.][]{hameury83,brown98,dipanjan12}. A fixed boundary, although preserving the physical nature of the magnetic field at the boundaries, suffers from the disadvantage that it cannot truly address the question of ``field burial", since the magnetic field at the boundaries cannot change from their initial value.  A possible way  to circumvent this numerical constraint is to  fix the external boundary to a dipole field with a variable dipole moment, whose value is evaluated by fitting the $\psi$ field below the boundary. Though attempted in \citet[][hereafter PM04]{melatos04}, it was not successfully implemented due to  reported numerical difficulties. For the results presented in Sec.~\ref{sec.results} we consider the fixed boundary condition with the outer boundary fixed to that of a dipole. 

\subsection{Specifying $\mathbf{r_0}(\boldsymbol{\psi})$ to define the mound shape}\label{sec.moundheight}
The arbitrary function $r_0(\psi)$ has been specified using two different prescriptions so far. One implementation assumes $r_0(\psi)$ to be a simple analytic function of $\psi$ with a well defined derivative \citep{hameury83,brown98,melatos01,dipanjan12}, e.g.
\begin{align}
r_0(\psi) &= R_s+r_c\left(1-\left(\frac{\psi}{\psi_a}\right)^2\right) \label{eq.parabolic} \\
\mbox{ or, } r_0(\psi) &= R_s+\frac{r_c}{0.25}\left(0.25-\left(\frac{\psi}{\psi_a} -0.5\right)^2\right), \label{eq.hollow}
\end{align}
 which can be readily applied to eq.~\ref{eq.gs}, as done for the results presented later in Sec.~\ref{sec.results}. Here $r_c$ is the maximum height of the mound and $\psi_a$ is the flux function at the edge of the polar cap. The extent of the polar cap is defined by the field line connecting the neutron star surface to the radius in the accretion disk plane. The Alfv\'en radius is given by \citep{mukhe15}
\begin{align}
r_A &= 3.53\times10^3 \mbox{ km}\left(\frac{B_s}{10^{12} \mbox{ G}}\right)^{4/7}\left(\frac{R_s}{10 \mbox{ km}}\right)^{12/7} \nonumber \\
& \times \left(\frac{\dot{M}}{10^{-9} M_\odot\mbox{ yr}^{-1}}\right)^{-2/7} \left(\frac{M}{1.4 M_\odot}\right)^{-1/7}. \label{eq.alfven}
\end{align}
This choice of the flux function is arbitrary, and different forms affect the shape and structure of the mound \cite{dipanjan12}. For example, eq.~\ref{eq.parabolic} defines a mound with its maximum height at the pole ($\theta=0$), as shown in Fig.~\ref{fig.gs}. However, a more realistic description is given by eq.~\ref{eq.hollow}, which defines a mound shaped as a ring, peaking at $\psi=0.5 \psi_a$ (see Fig.~\ref{fig.hollow}). A ring shaped accretion profile is expected for a mass loading at the accretion disk with a finite radial extent 
beyond the truncation radius \citep{ghosh78,ghosh79}.

A different approach has been pursued by \citet{melatos04,priymak11,priymak14}, where the flux function has been determined from the  mass distribution in flux tubes
\begin{equation}
\frac{dM(\psi)}{d\psi} = 2\pi \int_s ds \, r\sin\theta \rho (r,\psi) \frac{d\psi}{\nabla \psi} \label{eq.massflux}
\end{equation}
The integral is performed along a constant $\psi$ contour. For a classical isothermal equation of state, eq.~\ref{eq.massflux} can be readily inverted to express $r(\psi)$ as a simple analytic function of $dM/d\psi$, which is then subsequently specified, as in \cite{melatos04}. For a degenerate gas or a polytropic EOS, a simple analytic inversion cannot be obtained and $r_0(\psi)$ is determined iteratively starting from a guess value. Although more physically motivated, the choice of the function $dM/d\psi$, however, still remains arbitrary. 

PM04 assumes the mass profile to be 
\begin{equation}
M(\psi) = M_a\frac{\left(1-\exp(-\psi/\psi_a)\right)}{2\left(1-\exp(-\psi_*/\psi_a)\right)} \,,
\end{equation}
where $\psi_*$ is the flux function at the equator. However for such a mass profile only $\sim 63\%$ of the total mass is contained within the polar cap, with the rest being distributed over the field lines extending up to the equator. This implies significant mass-loading of field lines well inside the Alfv\'en radius. Although several works have shown that accretion disk may not strictly be truncated at the conventional Alfv\'en radius \citep{spruit90,sprui93,dange10,roman08}, the effect will be significant for only low field LMXBs with a dipole of strength $\sim 10^{25}$ G cm$^{3}$. For high field pulsars ($\mu \gtrsim 10^{29}$ G cm$^{3}$, as considered in PM04) with $r_A \sim 3000$ km, the excursion of the disk beyond the conventional Alfv\'en radius will be modest. 

Such a distribution is thus not consistent with the assumed initial magnetosphere model with $\mu \gtrsim 10^{30}$, as in PM04. Assuming a mass distribution extending to the equator results in building of very large ``mountains", which would otherwise be smaller if the mass is strictly confined to the polar cap. Ideally the choice of $dM/d\psi$ should be guided by the accretion profile arising from the disk-magnetosphere interaction \cite[e.g. in ][]{ghosh78,ghosh79}, which has not been self consistently modelled so far. Thus both methods pursued in the literature so far suffer to some extent from the arbitrariness of the choice for the shape of the mound.

\subsection{Local distortions in magnetic field topology}\label{sec.results}
\begin{figure*}
	\centering
        \includegraphics[width = 8cm,keepaspectratio] {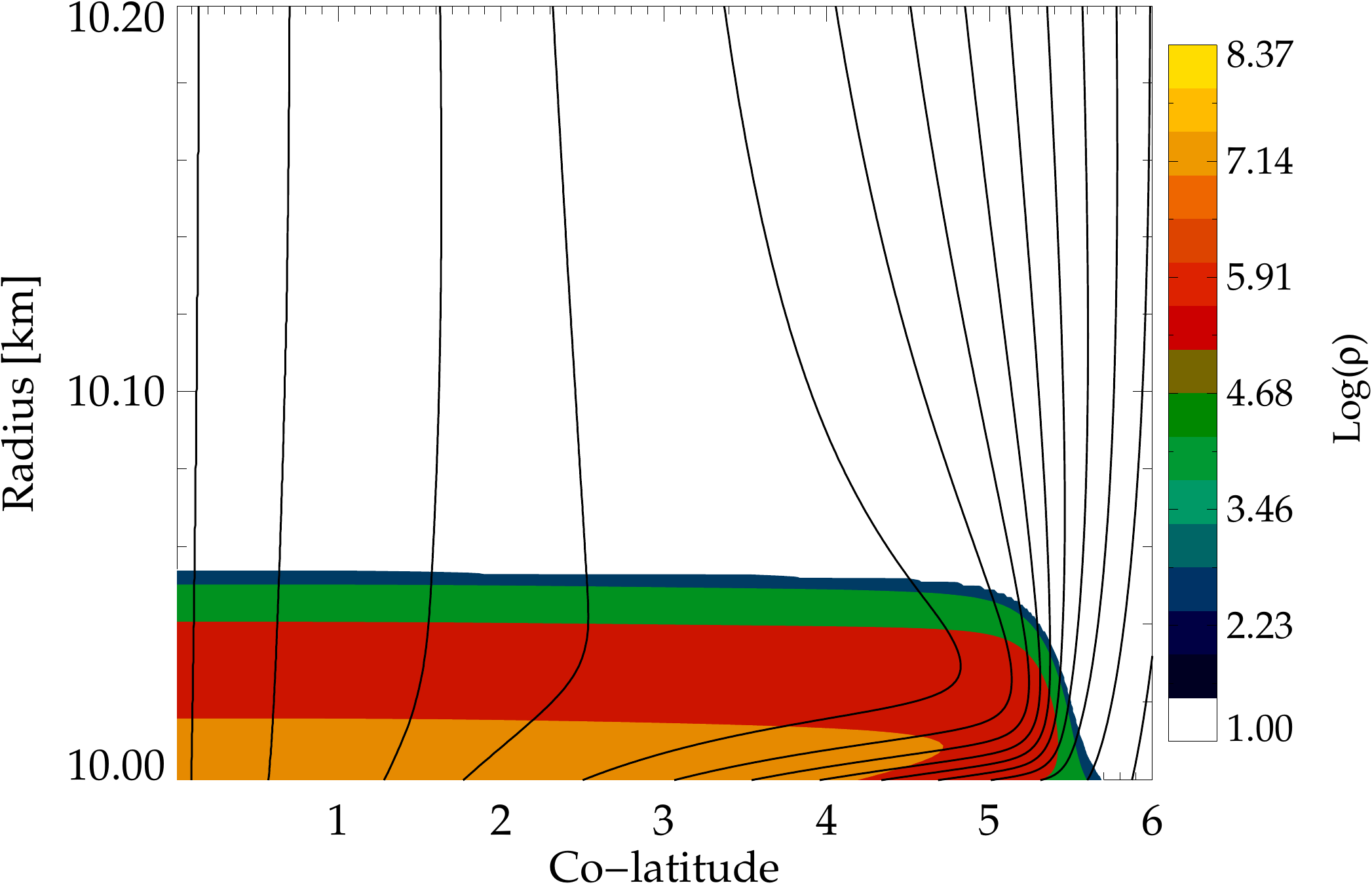}
        \includegraphics[width = 8cm,keepaspectratio] {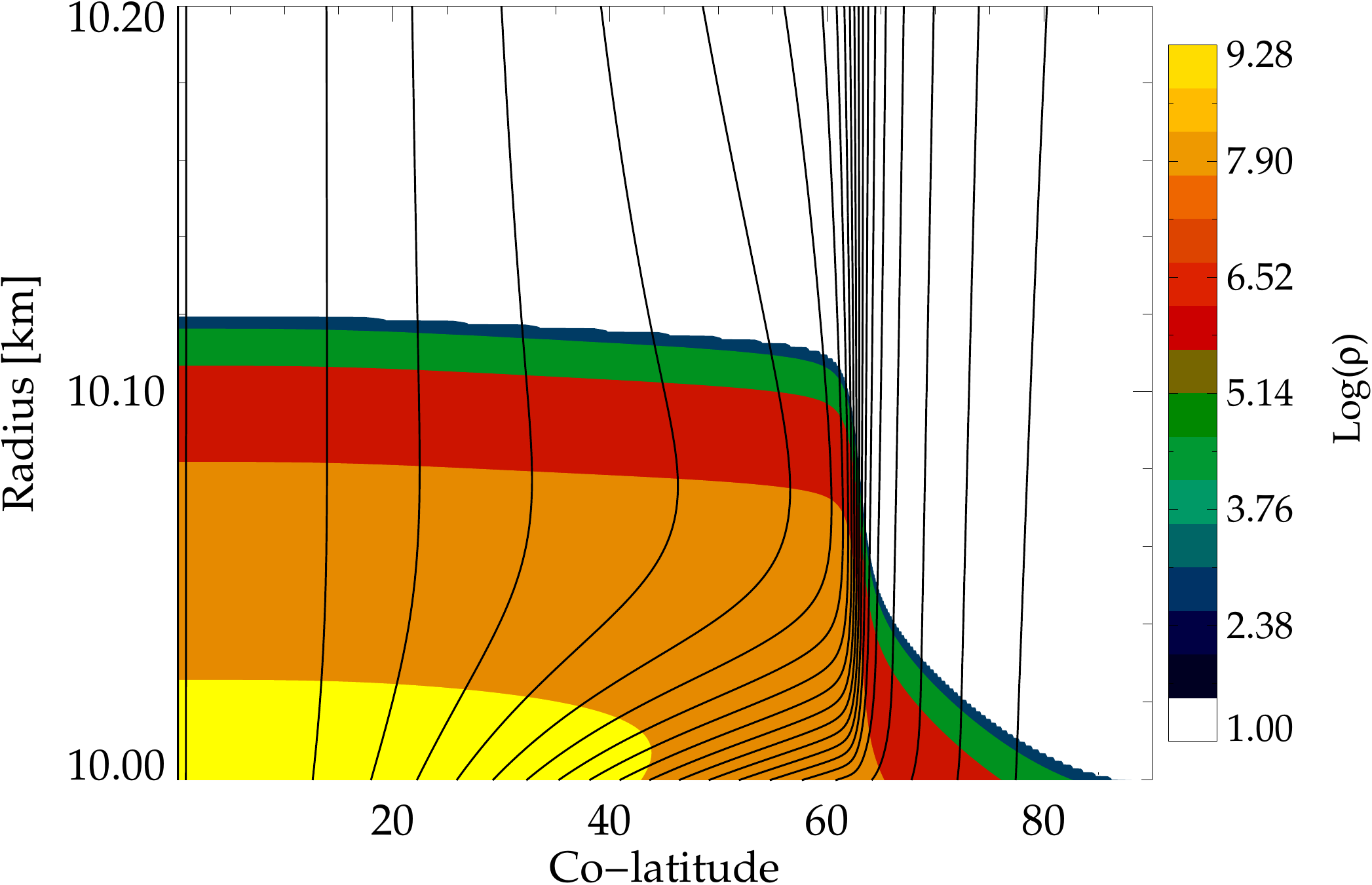}
	\caption{ \small Solutions to the GS equation with Paczynski EOS. \textbf{Left:} Field lines (in black) and density ($\log(\rho)$ in coloured contours) for a mound of central height $r_c=54$m, with a polar cap of $\sim 1$ km. The y-axis is the radius from the centre of the neutron star. The x-axis is the
co-latitude in degrees. Total mass is $\sim 4\times 10^{-12} M_\odot$. \textbf{Right:} A mound of central height $r_c=120$m, with matter distributed up to co-latitude $\sim 90^\circ$. Total mass enclosed is $\sim 8\times10^{-9} M_\odot$.  }\label{fig.gs}
\end{figure*}
\begin{figure}
	\centering
        \hspace{-1cm}\includegraphics[width = 7.9cm,keepaspectratio] {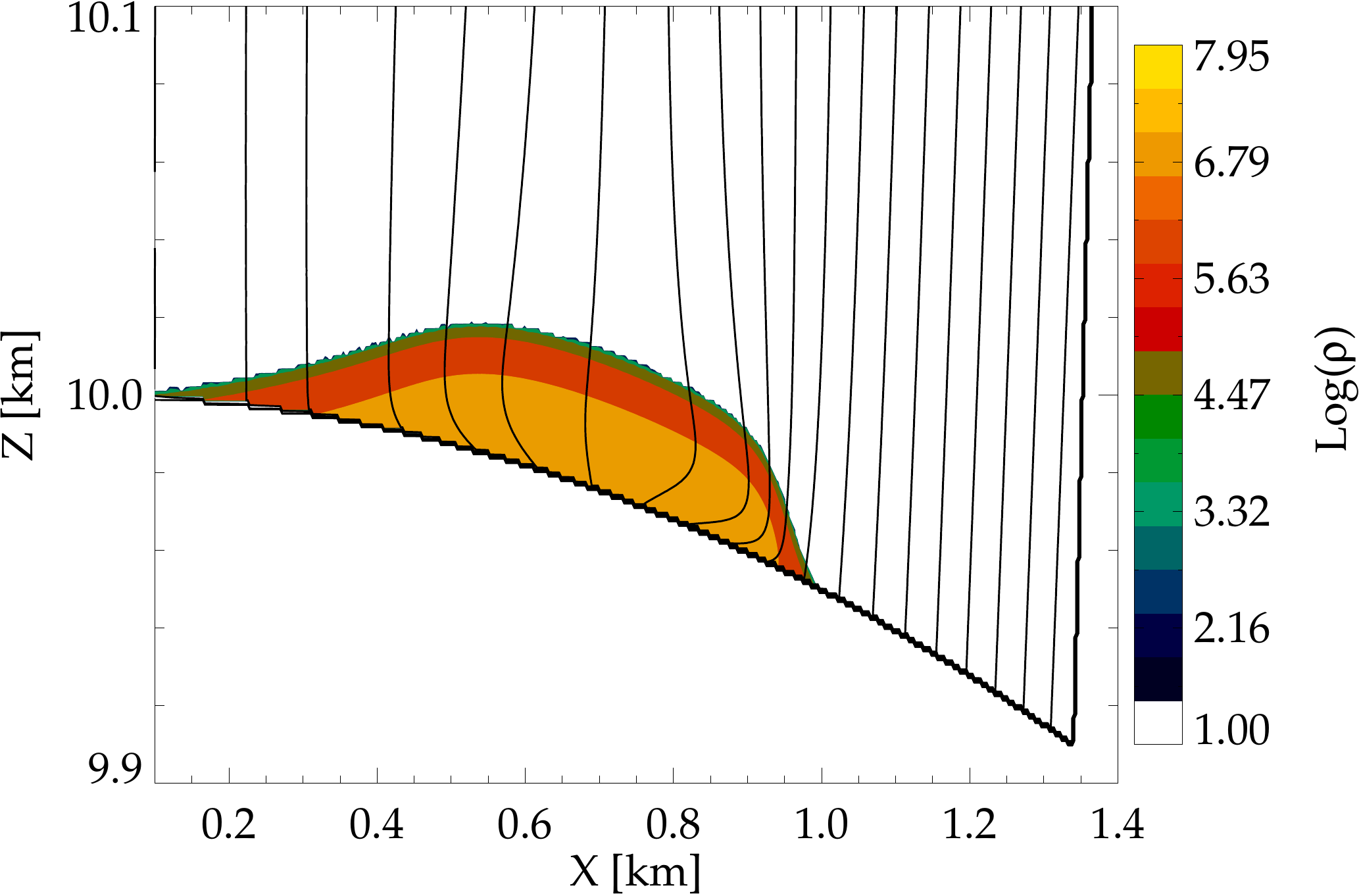}
	\caption{ \small A mound shaped as a ring with central height $r_c=36.5$m and total mass $\sim 9 \times 10^{-13} M_\odot$.  }\label{fig.hollow}
\end{figure}
In this section we present the solutions of the Grad-Shafranov equation with the Paczynski EOS (eq.~\ref{eq.pacz}). The solutions are obtained using an iterative numerical scheme as outlined in \citet[][hereafter MB12]{dipanjan12}. We consider the outer boundary to be fixed to that of a dipole field (as discussed in Sec.~\ref{sec.boundary}). The mound shape is defined by a simple analytic function (eq.~\ref{eq.parabolic} and eq.~\ref{eq.hollow}) as outlined in Sec.~\ref{sec.moundheight}. The initial starting guess solution is taken to be that of a dipole field:
\begin{align}
\psi _d &= \psi _E \frac{R_s}{r}\sin^2\theta, \\
\mbox{ where } \psi_E &= \frac{B_0 R_s^2}{2}. 
\end{align}
In the above, $\psi_E$ is the flux function at the equator where the surface magnetic field strength is $B_0$. Thus the flux function at the polar cap is $\psi _a = \psi_E \sin^2\theta _a$, where the $\theta_a$ defines the co-latitude of the polar cap edge. In Fig.~\ref{fig.gs} we present the solutions of the GS equation using the new Paczynski EOS (as derived in Sec.~\ref{sec.pacz} above). Near the base of the mound there is significant deviation of the field lines from the undisturbed dipole value. Lateral pressure from the confined matter is balanced by the tension arising from the curvature in the magnetic field lines.  The amount of distortion depends on the total mass enclosed. The largest distortions occur at the edge of the mound where the pressure gradients are highest.

There some notable differences as well as similarities to the results previously reported in \citet{melatos04,priymak11,dipanjan12}:
\begin{itemize}
\item\emph{Compact yet massive mounds:} 
\begin{figure}   
	\centering
        \hspace{2mm}\includegraphics[width = 7cm,keepaspectratio] {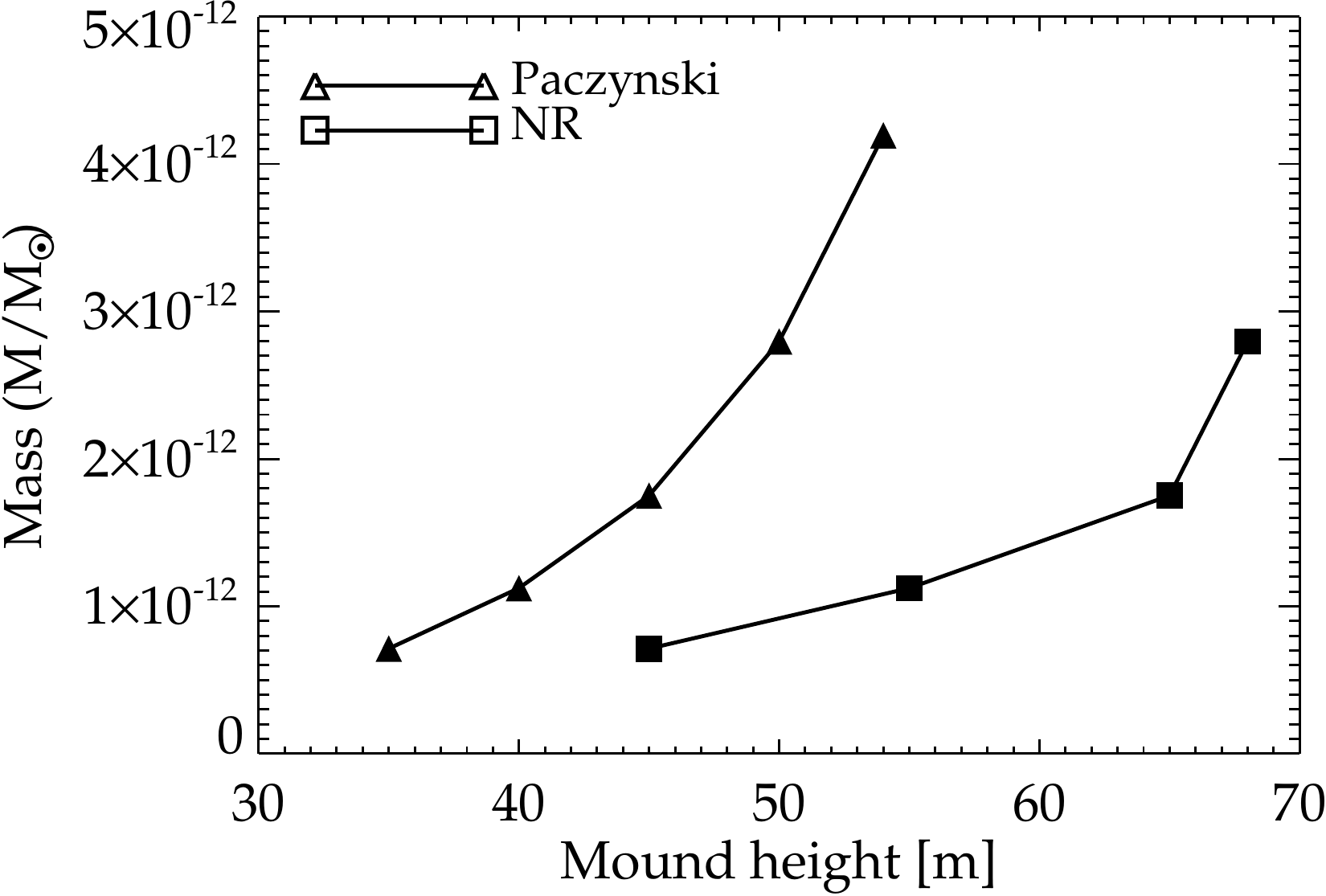}
	\caption{ \small Comparison of mass of confined mound as a function of mound height for two different plasma equation of states: a) non-relativistic (NR): $p\propto\rho^{5/3}$ (square) b) Paczynski EOS given by eq.~\ref{eq.pacz} (triangles). For a similar maximum central height, the Paczynski EOS confines larger mass. }\label{fig.solcompare}
\end{figure}
Compared to the non-relativistic EOS ($p\propto \rho^{5/3}$) used in MB12, the solutions with the Paczynski EOS can accomodate higher mass at a lower mound height, for a given initial surface field strength. A comparison of the mound masses for different mound heights ($r_c$) for solutions with the Paczynski and NR EOS is presented in Fig.~\ref{fig.solcompare}. Solutions with the Paczynski EOS show larger deformation of magnetic field lines for mounds of similar central height. Larger curvature in field lines results in larger mass contained in a flux tube, yielding more compact mounds of higher total mass as compared to the results in MB12. 

\item\emph{Large mountains:} The solutions in Fig.~\ref{fig.solcompare} are for mounds strictly confined within a polar cap of radius $\sim 1$ km ($\theta_a \sim 5.7^\circ$), roughly corresponding to an Alfv\'en radius (eq.\ref{eq.alfven}) for an Eddington accretion rate. Considering a mound of larger extent is inconsistent with the physics of magnetospheric interaction, as discussed in the earlier in section (Sec.~\ref{sec.pacz}). \citet{priymak14} point out that the difference in mound mass between MB12 and their work are primarily due to the difference in the way the mound height function is specified (outlined earlier in Sec.~\ref{sec.pacz}). 

However, here we note (as in Fig.~\ref{fig.gs}) that even with a simple analytic prescription of the mound height function, as in MB12 and this work, mounds of similar large masses can also be obtained, without resorting to preserving the mass per flux tube  (eq.~\ref{eq.massflux}), as done in PM04. If the accreted mass in not strictly confined to the polar cap, and distributed up to the equator, the total mound mass enclosed is similar to the mound sizes in \citet{priymak14} for the non-relativistic EOS. 

For example, for the solution presented in Fig.~\ref{fig.gs}, the total mass confined is $\sim 8 \times 10^{-9} M_\odot$, which is close to the maximum mass of numerically converged solutions for the model B (non-relativistic EOS) reported in \citet{priymak11}. Such a mass is 3 orders of magnitude (or more) larger than the mounds strictly confined within a polar cap of $\sim 1$ km radius (as in MB12 and the results in Fig.~\ref{fig.solcompare}). Hence large confined mountains can also be formed by the method outlined in MB12, although not fully consistent with the physically motivated model of mass loading from an accretion disk truncated at the Alfv\'en radius. 

\end{itemize}

\section{Instabilities}\label{sec.instabilities}
The Grad-Shafranov solutions discussed above do not guarantee stability to perturbations. If the solutions are unstable to MHD instabilities, then the formation of large scale confined mountains due to steady accretion will not be sustainable. The large curvature in the magnetic field lines makes them extremely susceptible to interchange mode instabilities, whose effect we discuss below.
\subsection{Solutions with closed magnetic loops}\label{sec.threshold}
A form of the instability arises during the course of the numerical iterative process of obtaining the Grad-Shafranov solutions, as discussed in \citet{hameury83,melatos04,dipanjan12}. Beyond a threshold mound height, field lines with closed magnetic loops appear within the domain, and the iterative scheme does not converge numerically to yield a solution. Beyond this limit, the curvature in the field lines becomes unable support the pressure gradients. \citet{mukhe14}, and also MB12, show that a simple scaling  relation can be derived from equating the pressure gradients to the magnetic curvature at the threshold height ($h_T$)
\begin{equation}
h_T \propto B_n^{4/9} R_p^{4/9},
\end{equation}
where $B_n$ is the normal component of the magnetic field at the base and $R_p$ the polar cap radius. This matches well with the maximum height for the parabolic profile (eq.~\ref{eq.parabolic}), beyond which numerical solutions fail \citep[as shown in Sec. 2 of ][]{mukhe14}. The failure of convergence is not just a deficiency of the numerical scheme, but an inherent topological problem where closed magnetic loops are allowed in the solution domain, as discussed in PM04. In reality, such closed magnetic loops would form buoyant bubbles which will rise away from the surface. 

\subsection{Pressure driven MHD instabilities}\label{sec.litwin} 
The ballooning mode is source of pressure driven instability commonly encountered  in magnetically
confined plasma with $\beta > 1$\footnote{Plasma $\beta$ is the ratio of gas to magnetic pressure: $\beta=P_{\rm gas}/P_{\rm magnetic }$.}, such as in tokamaks \citep{friedberg82}. Such instabilities are also
applicable to magnetically confined mounds on neutron stars \citep{litwin01}, where pressure 
from the dense degenerate plasma may dominate over the magnetic pressure. From the energy intergal of 
the perturbed mound, as outlined in  \citet[][]{litwin01} and \citet[][see eq. 15]{dipanjan13a}, instabilities
in the magnetically confined mound can arise from an interplay between the curvature of the field line 
($(\hat{\mathbf{b}}.\nabla)\hat{\mathbf{b}}$) and the pressure gradient. From linear stability analysis, \citet{litwin01}
had identified that such instabilities will set in for significant local distortion of the dipole magnetic field, with  $B_r/B_z \gtrsim 11.7$ \citep[eq. 55-57 and eq. 62 of ][]{litwin01}. The onset of the instability is predicted to occur for an accreted mass 
\begin{align}
M &= 1.6\times10^{-12} M_\odot \left(\frac{B}{10^{12} \mbox{ G}}\right)^2 \left(\frac{R_p}{1 \mbox{ km}}\right)^3 \nonumber \\
  & \times \left(\frac{\mu_e}{2}\right)^{4/3} \times \left(\frac{\rho_b}{10^8 \gcc}\right)^{-1/3}  \label{eq.litwin}
\end{align}
\citep[following eq. 63 of][]{litwin01}, much smaller than the large mountains required for appreciable effects of field burial found in PM04. In eq.~\ref{eq.litwin} above, $\rho _b$ is the mean density at the bottom of the mound. An ultra-relativistic EOS has been assumed \citep[as done in][]{litwin01}. We assume $\mu_e=2$ to be consistent with the results presented in this work.

\subsection{Numerical simulations}
Presence of ballooning mode instabilities were numerically verified using MHD simulations of the perturbed
Grad-Shafranov solutions \citep{dipanjan13a,mukhe13d}. 2D simulations \citep{dipanjan13a} confirm the  threshold
discussed earlier in Sec.~\ref{sec.threshold}, where addition of mass beyond the threshold mound height destabilises
the equilibrium. This leads to formation of closed magnetic loops triggered by magnetic reconnection of the in falling, unsupported matter. 

Pressure driven ballooning modes arise in 3D simulations \citep{mukhe13d} from the growth of perturbations imposed on the Grad-Shafranov solutions. The 3D simulations identify a stability threshold close to that predicted from the analytical linear analysis in \citet{litwin01}. Previous MHD simulations \citep{payne07,vigelius08} did not find the presence of such instabilities, from which they concluded the solutions to be stable to perturbations, in contention with the predictions of \citet{litwin01}. The absence of the instabilities in these works is likely due to insufficient spatial resolutions required to resolve the unstable modes. \citet{mukhe13d} found that spatial resolution of $\delta x \leq 1$ m is required to fully capture the growth of the instabilities. Lower resolution results in higher magnetic diffusivity, allowing the perturbations to dissipate with relative ease without affecting the equilibrium. 

\subsection{Effect of non uniform composition and thermal stratification}\label{sec.otherinst}
The results presented in this and previous works on magnetically confined accretion mounds have assumed a plasma with uniform composition and temperature. However, such an assumption is not strictly true when fresh accreted matter arrives on the neutron star's surface during an active mass-transfer phase from the binary companion. The infalling matter undergoes a series of nuclear reactions starting from the upper atmosphere ($\rho \sim 10^5-10^6 \gcc$) where infalling hydrogen and helium is burnt to carbon and higher elements \citep{schatz99,bilds98,brown98}. Electron capture processes can operate at higher densities $\gtrsim 10^7 \gcc$ yielding heavier elements \citep{bilds98}. The burnt ashes are finally compressed to deeper layers as they settle under the influence of gravity.

The energy released from the nuclear reactions, as well as the gravitational settling, results in thermal and chemical stratification of the settling layer. Such stratification can potentially drive buoyancy driven g-mode instabilities \citep{bilds98,cumming01}. \citet{cumming01} have shown that although at high accretion rates ($\dot{m} >0.02 \dot{m}_{\rm Edd}$) magnetic field may be buried and compressed to deeper layers, buoyancy driven instabilities may operate beyond a threshold magnetic field of $B_c\sim 10^{10}-10^{11} \mbox{G}$ in the ocean. This will limit the efficiency of field burial process. However, the above works have assumed a simple plane-parallel geometry with horizontally stratified magnetic field. A self consistent modelling of the magnetic field configuration of a settling flow in the neutron star ocean has not been carried out so far.

\section{Summary and Implications for field burial}\label{sec.summary}
In the sections above we have reviewed the existing work on modelling the magnetic confinement of accreted matter on a neutron star and presented new results with a more accurate equation of state of the confined degenerate plasma. For mounds strictly confined to a polar cap of radius $\sim 1$ km, on a neutron star with surface  magnetic field of $\sim 10^{12}$ G, the Grad-Shafranov equation can be solved for the magnetic equilibria for mounds of masses up to $\sim 10^{-12} M_\odot$. For larger mounds, numerically converged solutions to the Grad-Shafranov cannot be obtained.

A larger mass ($\sim 10^{-8} M_\odot$) can be accommodated if the matter is not strictly confined to the polar cap, but distributed  all the way up to the equator, as in \citet{priymak11}. Such a distribution, however, is inconsistent with the expected scenario of magnetic channelling of accretion flow from a truncated accretion disc at or near the Alfv\'en radius. Solutions for even larger mass $\sim 10^{-4} M_\odot$ have been obtained by \citet{melatos04} using an isothermal equation of state for the plasma. However, assuming an isothermal plasma is incorrect for the densities inside the mound where matter will be degenerate.

The magnetic fields of accreting millisecond pulsars lie within the range of $10^7-10^9$ G \citep[$\mu \sim 5\times10^{24}-10^{26}$ G cm$^{-3}$,][]{mukhe15}. With the largest accretion mounds allowed by the Grad-Shafranov modelling \citep{melatos04,priymak11}, the dipole moment was found to reduce by about $\sim 10^{-2}$ times the initial value. A further reduction of the dipole moment by another 2 orders of magnitude from the starting value of $\mu \sim 5\times10^{29}$ G cm$^{-3}$ ($B\sim 10^{12}$ G) is required to explain the field strengths of MSPs. Relative reduction of dipole moment by a factor of $10^{-3}$ has been achieved by a bootstrap accretion method in \citet{payne07}. However, the work still suffers from the use of the unphysical isothermal equation of state which accommodates larger mass due to lower gas pressure. For the models following a strict confinement of the accreted matter in the polar cap, the mound is much lower ($\sim 10^{-12} M_\odot$), which will have even less effect on the apparent reduction of the dipole moment. The original suggestions of the field  burial process \citep{romani90,cumming01} involved simplified model of field geometry. However, it appears that following a more physically motivated model of magnetic confinement, the efficiency of the burial process is not sufficient to explain the reduction of the field strength from $10^{12}$ G to $10^{8}$ G. 

Formation of very large confined mountains will also be limited by MHD instabilities. Such instabilities have been shown to operate for mound masses larger than $\sim 10^{-13} M_\odot$ \citep{litwin01,dipanjan12,mukhe13d}, much lower than the masses required for any appreciable reduction in dipole moment. When masses greater than the threshold is accreted, MHD instabilities will set in, transporting the settling matter across magnetic field lines, without significantly distorting them. If the large scale global distortions are restricted by MHD instabilties, the efficiency of field burial process will be significantly reduced.   Thus it appears that reduction of dipole moment by burial of magnetic field with accreted matter is not an effective mechanism. Other proposed methods of field reduction, namely accretion induced enhancement of ohmic decay \citep{konar97a,konar99a,konar99b} are more promising alternatives to address the origin of reduced magnetic field of millisecond pulsars, and the connection to the accretion history of the neutron star. 

\section{Future directions}\label{sec.future}
Although the arguments presented above are pessimistic towards the field burial scenario, there exists significant scope of improving the existing works. Firstly the several drawbacks of the Grad-Shafranov solution with regards to handling the boundary condition and maintaining a physically consistent framework needs to be addressed. Secondly, although \citet{dipanjan13a,mukhe13d} have confirmed the presence of MHD instabilities predicted from linear analysis \citep{litwin01}, the works do not address how the matter spreads out of the polar cap. The simulations performed have been restricted to within the body of the mound due to numerical limitations. Although the interchange instabilities may lead to leakage of matter, how this settles outside the polar cap is yet to be studied. 

The works on accretion induced ohmic decay \citep{konar97a,konar99a,konar99b} also suffer from restrictive assumptions of spherical symmetry and do not account for a physically motivated geometry of an accretion flow. \citet{choudhuri02} and \citet{konar04} addresses this to some extent by evaluating the field evolution subject to an imposed flow pattern. However, the works do not self consistently model the impact of the gas and magnetic pressures in determining the evolution of the field topology. Although global simulations of magnetospheric accretion have been addressed in recent years \citep{romanova02,roman04,roman08}, self consistently modelling the accretion physics with sufficient resolution to track the surface field deformation is computationally challenging. 

Future works need to explore the dynamic evolution of the spread of matter from the mound, its thermal structure and implications for long term evolution of the magnetic field. Recent observations of the time evolution of cyclotron resonant scattering features such as in Her X-1 \citep{staub16} point towards accretion induced deformation of the polar cap magnetic field being observed over a span of a few decades. This implies a short term deformation of the magnetic field before the spread from the polar cap, whose imprint might be an increase in the hot spot surface area \cite[e.g. as conjectured in ][]{ferrigno13}. Thus to conclude, the mechanism by which accretion affects the magnetic field evolution on neutron stars is still ill-understood, and requires better physically motivated models.

\section*{Acknowledgement}
Majority of the results discussed in this review are from work performed as part of my PhD thesis under the supervision of Dr. Dipankar Bhattacharya. I have greatly benefitted from the exchange of ideas and collaborations with several colleagues on topics related to magnetic field evolution on neutron stars, and gratefully thank Dipankar Bhattacharya, Sushan Konar, Henk Spruit, Andrea Mignone, Michiel van der Klis, Yuri Levin and Andrew Melatos for fruitful discussions. I also thank the referee for the thorough scrutiny and helpful suggestions to improve the text.

\def\apj{ApJ}%
\def\mnras{MNRAS}%
\def\aap{A\&A}%
\def\apjl{ApJ}
\def\physrep{PhR}
\def\apjs{ApJS}
\def\pasa{PASA}
\def\pasj{PASJ}
\def\nat{Nature}
\def\memsai{MmSAI}
\def\aj{AJ}%
\def\aaps{A\&AS}%
\def\iaucirc{IAU~Circ.}%
\def\sovast{Soviet~Ast.}%
\def\apss{Ap\&SS}

\bibliography{mnrasmnemonic,dip_1.5.15,dipanjanbib}
\bibliographystyle{mnras}

\end{document}